\documentclass[prb,preprint]{revtex4}
\usepackage{times}
\usepackage{bm}
\usepackage{graphicx}
\usepackage{fancyhdr}
\usepackage{ae}
\begin{document}
\date{\today}
\title{Potential of mean force and the charge reversal of rodlike polyions}
\author{J\"urgen F. Stilck\\
Instituto de F\'{\i}sica - UFF\\
Av. Litor\^anea s/n\\
24.210-340 - Niter\'oi - RJ\\
and
Yan Levin\\
Instituto de F\'{\i}sica - UFRGS, \\
Caixa Postal 15051, CEP 91501-970,\\Porto Alegre,
Brazil}

\begin{abstract}

A simple model is presented to calculate the potential of mean
force between a polyion and a multivalent counterion 
inside a polyelectrolyte solution. 
We find that under certain conditions the
electrostatic interactions can lead to a strong attraction
between the polyions and the multivalent counterions, favoring
formation of overcharged polyion-counterion complexes. 
It is found that small concentrations of salt enhance the overcharging 
while an excessive amount of salt hinders the charge reversal.
The kinetic limitations to overcharging are also examined.

\end{abstract}

\maketitle

\section{Introduction}

It is our pleasure to contribute this paper to the special
issue of Molecular Physics dedicated to celebrate 
Ben Widom's outstanding
contributions to Physical Chemistry and Statistical Mechanics.  Ben's
work is characterized by a profound physical insight, combined
with an ability to abstract the most complex physical phenomena 
into a simple model.  From  scaling and criticality~\cite{Wi65} 
to microemulsions~\cite{Wi84} and the hydrophobic effect~\cite{Wi99}, Ben's 
sagacity has  opened  new frontiers of 
Physical Chemistry.  While it is impossible to compete with Ben's
intuition, one can at least {\it try} to follow his 
example. In this paper we will, therefore, study a simple model
of interaction between a polyion and multivalent counterions
inside a polyelectrolyte solution.  

Thermodynamic systems in which long range Coulomb interactions play the
dominant role pose an outstanding challenge to 
Physical Chemistry~\cite{Le02}.
Even such basic question as the possible existence of a liquid-gas  phase
separation in a restricted primitive model has been positively settled
only quite recently~\cite{Le02}.  
Even so, the order of this transition still
remains a source of an outstanding debate~\cite{LuFiPa01}.  
For strongly asymmetric
electrolytes such as aqueous colloidal suspensions, even the 
existence of a liquid-liquid phase separation continues to be 
controversial~\cite{RoHa97,LeBaTa98,DiBaLe01,DeGr02,TaSc03,TrLe04}.  

When aqueous colloidal suspensions or 
polyelectrolyte solutions contain multivalent 
counterions other curious
phenomena appear.  For example, it is 
found that for  sufficiently
small separations 
two like-charged polyions can attract one 
another~\cite{RoBl96,GrMaBr97,HaLi97,ArStLe99,HaLo00,GeBrPi00,SoCr01,AnLiWr03}. 
If an external electric field is applied to such a suspension 
the electrophoretic mobility of colloidal particles
is often found to be reversed, so that the particles
move in the direction opposite to the one expected 
based purely on their chemical
charge~\cite{LoSaHe82,MeToLo01,GrNgSh02,Le02,MaQuGa03}.  
Both of these phenomena are a consequence of strong
electrostatic coupling between the polyions and the counterions.  

\begin{figure}
\begin{center}
\includegraphics[scale=1]{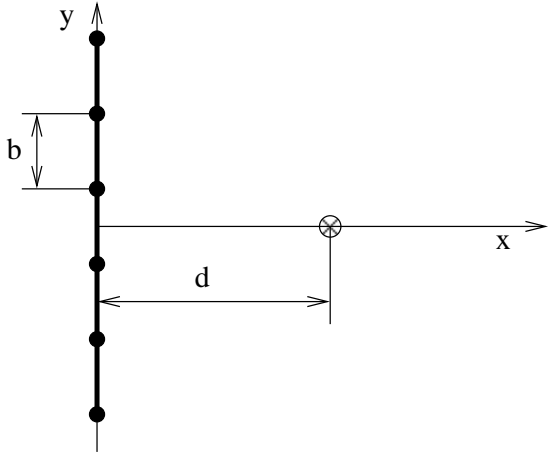}
\caption{Rodlike polyion with $Z=6$ monomers separated by a distance 
$b$ and a counterion located at $\rm x=d$.}
\label{conf}
\end{center}
\end{figure}

The counterions inside the suspension can be divided into
two categories: those which are associated (condensed) 
to the colloidal particle and those which are free.  The condensed
counterions contribute to the effective, renormalized, 
charge of the polyion-counterion
complex, while the free counterions and coions result in
screening of the electrostatic interactions inside 
the suspension~\cite{Le02}.
In this paper we will explore the potential of mean force
between a rodlike polyion with $n$ associated
counterions and a counterion located at a 
transverse distance $d$ from the polyion center, Fig. \ref{conf}.

\section{The model}

Consider a 
rodlike polyion of $Z$ (even) monomers, each carrying a charge $-q$, inside 
an aqueous  suspension containing multivalent counterions and
salt.  The monomers
are located uniformly with separation $b$ along the rod. 
Strong electrostatic coupling between the
polyion and the counterions results in a condensation of 
$n$  $\alpha$-valent counterions onto the polyion. 
The condensed counterions are free
to hop between the monomers  of the polyion~\cite{ArStLe99}.  
If a monomer has an associated counterion, its charge is
renormalized to $(\alpha-1)q$. The free, uncondensed, counterions
and coions  
screen the electrostatic interactions, changing the potential between
the two charges $q_1$ and $q_2$
from the Coulomb to the Debye-Hückel~\cite{DeHu23} form
\begin{equation}
V(r)=\frac{1}{\epsilon} \frac{q_1 q_2 \exp(-\kappa r)}{r},
\label{1}
\end{equation}
where $\epsilon$ is the dielectric constant of the solvent and $\kappa$ 
is the inverse Debye length.
The question that we would like to address in this paper is what is
the potential of mean force between the polyion-counterion complex 
containing $n$ condensed $\alpha$-ions 
and an additional $\alpha$-valent counterion 
located transversely at distance $d$
from the polyion center, see Fig. 1.

To proceed, we assign to each monomer 
$i$ a lattice-gas 
variable $\sigma_i$, such that
$\sigma_i$ is equal to $1$ if a counterion is condensed onto cite $i$ 
and $0$ otherwise. 
For a given configuration $\{\sigma\}$, the interaction Hamiltonian 
between the complex and a counterion located at a transverse 
distance $d$ 
from its center is
\begin{eqnarray}
{\cal H} &=& \frac{1}{D} \sum_{i=1}^Z \frac{\alpha q^2(\sigma_i
\alpha-1)}{\sqrt{r_i^2+d^2}}\exp(-\kappa \sqrt{r_i^2+d^2}) \nonumber \\
& & +\frac{1}{2D} \sum_{i,i^\prime=1, i \neq
i^\prime}^Z \frac{q^2(\sigma_i
\alpha-1)(\sigma_{i^\prime}\alpha-1)}{b|i-i^\prime|} \exp(-\kappa b
|i-i^\prime|), 
\label{2}
\end{eqnarray}
where 
\[
r_i=\frac{2i-1-Z}{2}b.
\]
It is convenient to define the reduced distance 
between the polyion and the counterion  $x=d/b$,  the
reduced inverse Debye length $k=\kappa b$, and the 
Manning parameter~\cite{Ma69,Ma78} 
as $\xi=q^2/\epsilon k_ BTb$.  In terms of these
adimensional variables the reduced Hamiltonian, 
$H \equiv \beta {\cal H}/\xi$, becomes 
\begin{eqnarray}
H &=& \sum_{i=1}^Z(\sigma_i \alpha-1)\left[
\frac{2\alpha}{\sqrt{(2i-1-Z)^2+4x^2}}\exp(-k \sqrt{(2i-1-Z)^2+4x^2})
\right.\nonumber \\ 
& & \left. +\frac{1}{2}\sum_{i^\prime=1,i \neq
i^\prime}^Z \frac{\sigma_{i^\prime}\alpha-1}{|i-i^\prime|}
\exp(-k |i-i^\prime|)\right].
\end{eqnarray}

The partition function is a trace over all possible distributions
of  $n$ condensed counterions among the $Z$ polyion sites.  
There is a total of 
\[
N_c=\frac{Z!}{(Z-n)!n!}
\]
such configurations.  The partition function is then 

\[
Q={\sum_{\{\sigma\}}}^\prime \exp[-\xi H],
\]
where the sum is over the $N_c$ configurations $\{\sigma\}$ which obey
the constraint $\sum_{i=1}^Z \sigma_i=n$, denoted by the prime. It is
convenient to order the terms in the Hamiltonian by the distances between
the pair of interacting charges. This results in
\[
H=\sum_{i=1}^{Z/2} 2\alpha[(\sigma_i+\sigma_{Z-i+1})\alpha-2]
\left[\frac{\exp(-k\sqrt{(2i-1-Z)^2+4x^2)}}{\sqrt{(2i-1-Z)^2+4x^2}}\right]
\]
\[
+\sum_{j=1}^{Z-1}\sum_{i=1}^{Z-j}(\sigma_i \alpha-1)(\sigma_{i+j}\alpha-1)
\frac{\exp(-kj)}{j}.
\]
If we now define the Boltzmann factors
\[
x_j=\exp\left[\frac{-\xi\exp(-k\sqrt{(2j-1-Z)^2+4x^2})}
{\sqrt{(2j-1-Z)^2+4x^2}}\right],
\]
and
\[
y_j=\exp\left[\frac{-\xi\exp(-k j)}{j}\right],
\]
the contribution of each configuration to the partition function
will be a product of these factors raised to exponents which are
polynomials in $\alpha$, that is
\begin{equation}
Q=\sum_{i=1}^{N_c} \prod_{j=1}^{Z/2}x_j^{v_{i,j}}\prod_{j=1}^{Z-1}
y_j^{u_{i,j}}.
\label{2a}
\end{equation}
The polynomials
$v_{i,j}=-a_{i,j}\alpha+b_{i,j}\alpha^2$ and
$u_{i,j}=c_{i,j}-d_{i,j}\alpha+e_{i,j}\alpha^2$, have integer
non-negative coefficients.
The advantage of the simple model constructed above is that
for not too large values of $Z$ and $n$ the partition function
can be evaluated exactly with a help of a computer.

The potential 
of mean force (measured in units of $q^2/\epsilon b$)
between a polyion-counterion complex and an $\alpha$-ion located at $x$ is, 
\begin{equation}
\phi(\xi,k,\alpha,x)=-\frac{1}{\xi} \ln \frac{Q(x)}{Q(\infty)}.
\end{equation}
The potential is normalized so that $\phi(\infty)=0$.

The computer code which generates the 
partition function for given values of $Z$
and $n$ determines the set of integer coefficients of the
polynomials defined following the Eq.(\ref{2a}).
Each set of polynomial coefficients may 
correspond to more than one internal configuration of
the polyion, so that the degeneracy must also be taken into account.
All the data is stored on the computer and 
used to perform a
floating point calculation of the free energy. 

\section{Results and Discussion}

\begin{center}
\begin{figure}
\includegraphics[scale=0.7]{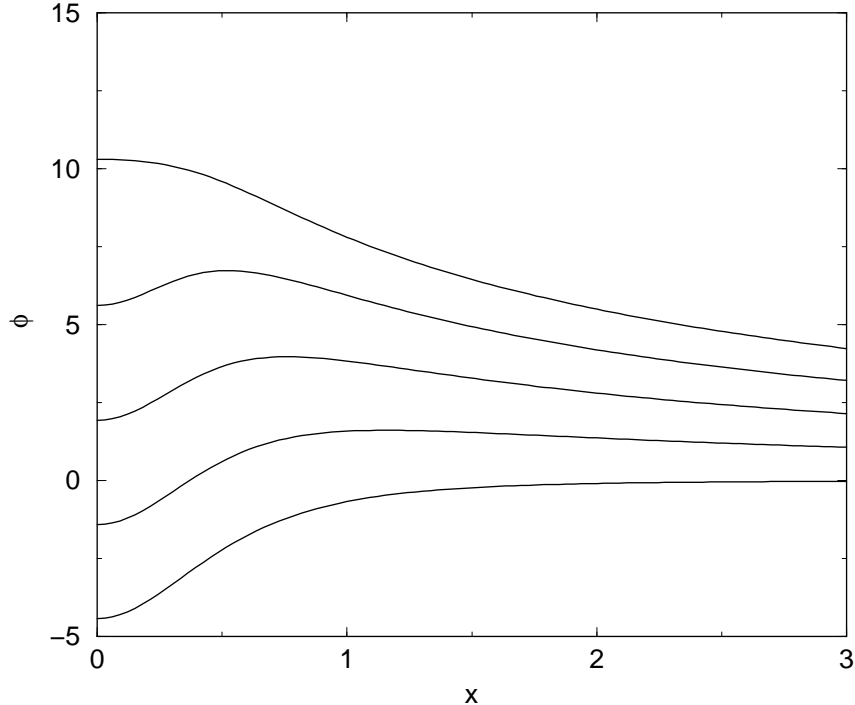}
\caption{Potential of mean force 
as a function of $x$ for $Z=10$, $\xi=1$, $k=0$,
  and $\alpha=2$. In upward order, the curves correspond to $n=5, 6,
  7, 8, 9$ condensed counterions.}
\label{phix}
\end{figure}
\end{center}

In Fig. \ref{phix} the potential of mean force between  
various complexes and an $\alpha$-ion is plotted.  
The complexes are composed of a polyion of charge $-10q$ and 
$n$ associated divalent
counterions.  Notice that for $n=5$  (neutral complex) the potential
is a monotonically increasing function of $x$, so that the sixth 
counterion is
always attracted to the complex.  
For an overcharged complex with $n=6$ condensed counterions, the potential of mean
force develops a barrier.  At large
distances the seventh counterion is repelled from the complex, while
at short distances it is attracted to it.  The minimum of the free energy,
however, is reached when the seventh counterion is located at
$x=0$. The potential of mean force, 
therefore, favors counterion condensation.  
The size of the barrier increases with $n$ and the minimum
at $x=0$ becomes metastable for 
$n=8$.  For $n=9$ the potential is a monotonically decreasing function of
$x$, and the tenth counterion is always repelled from the complex. 
We next study the dependence of the depth of the potential well  
and the height of the barrier on the parameters
of the model.
\begin{center}
\begin{figure}
\includegraphics[scale=0.7]{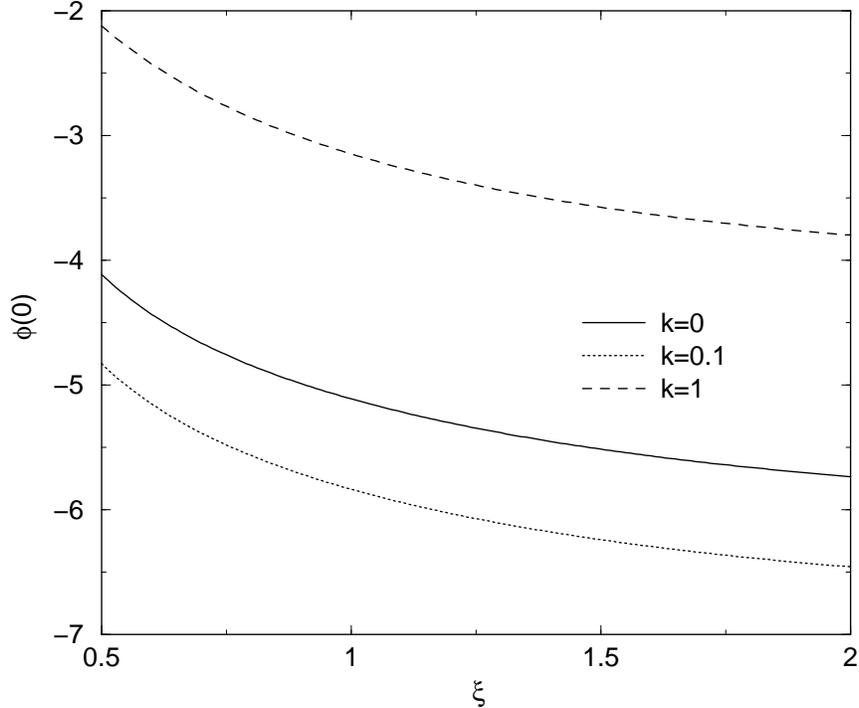}
\caption{The potential of mean force at $x=0$  as a
  function of $\xi$ for some values of $k$. The curves are for $Z=10$,
  $n=4$, and $\alpha=3$.}
\label{dphixi410}
\end{figure}
\end{center}

\begin{center}
\begin{figure}
\includegraphics[scale=0.7]{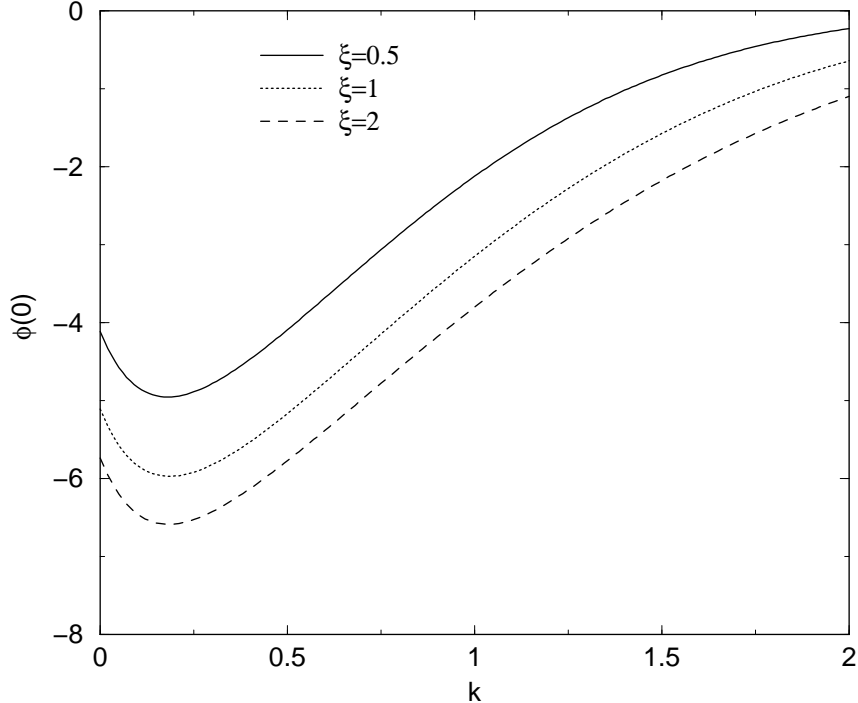}
\caption{$\phi(0)$ as a
  function of $k$ for some values of $\xi$. The curves are for $Z=10$,
  $n=4$, and $\alpha=3$.}
\label{dphik410}
\end{figure}
\end{center}

In Fig. \ref{phix},
we saw that when the complex is overcharged $n>Z/\alpha$, the potential 
can have two minima, one located at $x=\infty$ and another
at $x=0$.  Which one of the two minima is the global one
is determined by the sign of $\phi(0)$.  
Figs. \ref{dphixi410} and \ref{dphik410}  show the behavior of 
$\phi(0)$ as a function of $\xi$ and $k$.
When $\phi(0)<0$ the
position at $x=0$ is the absolute minimum, while when  $\phi(0)>0$,
$x=0$ is at most metastable. We should note, however, that the present
discussion is not sufficient to define the absolute number of condensed
counterions.  For a counterion to be condensed the depth of the potential
well must be sufficiently large, compared to the thermal energy $k_B T$, to
prevent its rapid escape from the polyion surface. 
At the level of the present discussion this criterion is arbitrary.  Thus,
in this paper we will not consider the absolute number of condensed
counterions but only the conditions which favor or disfavor the counterions
condensation.  From  
Figs. \ref{dphixi410} and \ref{dphik410}, 
we see that for a polyion of $Z=10$ and $n=4$
condensed trivalent counterions the minimum at $x=0$ 
is the global one for the parameters plotted.  Approach
of an additional fifth counterion to this already overcharged complex 
is, therefore, energetically favorable. 

\begin{center}
\begin{figure}
\includegraphics[scale=0.7]{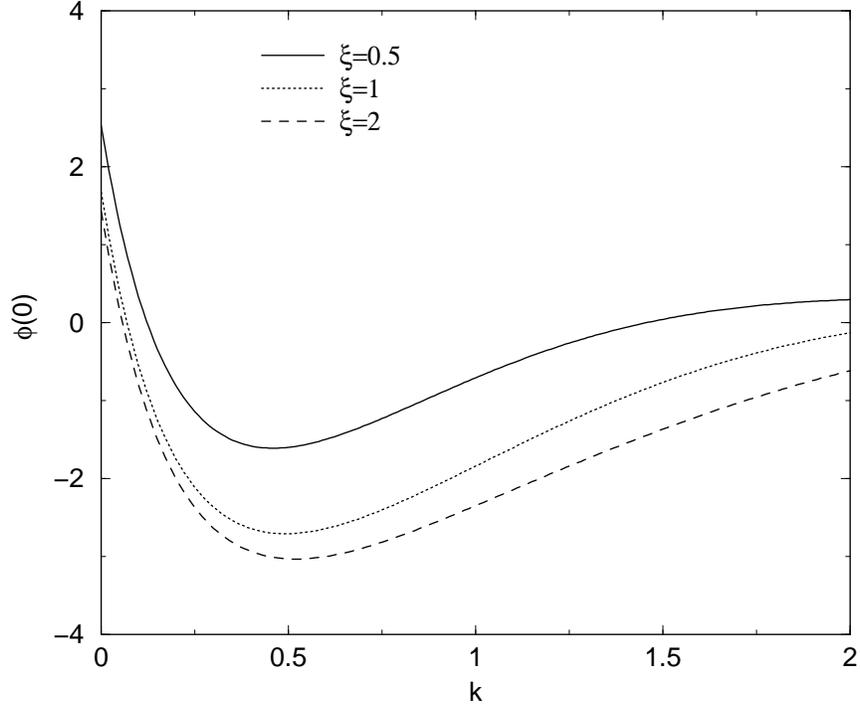}
\caption{$\phi(0)$ as a
  function of $k$ for some values of $\xi$. The curves are for $Z=10$,
  $n=5$, and $\alpha=3$.}
\label{dphik510}
\end{figure}
\end{center}

The depth of the global minimum $|\phi(0)|$ is a monotonically increasing
function of the Manning parameter, see Fig. \ref{dphixi410}.
The dependence on the salt concentration, however, is not
monotonic.  From Fig. \ref{dphik410} we see that small concentrations
of salt favor counterion condensation, i.e. $\phi(0)$ becomes
more negative for small $k$.  Larger concentrations
of salt, however, have a destabilizing effect on the counterion condensation.
This is even clearer for complexes composed of a polyion 
with $Z=10$ and 
$n=5$ condensed trivalent counterions.  
Fig. \ref{dphik510} shows that the position of the free
energy minimum is a non-trivial function of
salt concentration. Depending on the
Manning parameter $\xi$  and the concentration of salt $k$,
association of an additional, sixth, counterion can be 
either favored or disfavored.
On the other hand, for $Z=10$, $n=6$, and $\alpha=3$,   $\phi(0)$ 
is always positive so that a complex  
with $n=7$ condensed counterions can be at most metastable. 

\begin{center}
\begin{figure}
\includegraphics[scale=0.7]{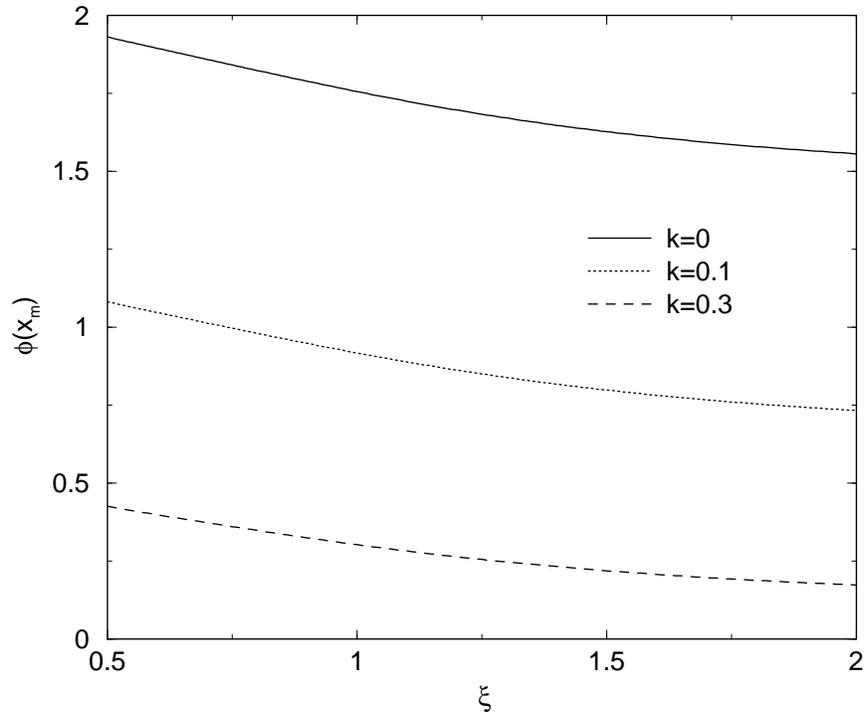}
\caption{The  barrier height  $\phi(x_m)$ as a
  function of $\xi$ for some values of $k$. The curves are for $Z=10$,
  $n=4$, and $\alpha=3$.}
\label{Dphixi410}
\end{figure}
\end{center}

\begin{center}
\begin{figure}
\includegraphics[scale=0.7]{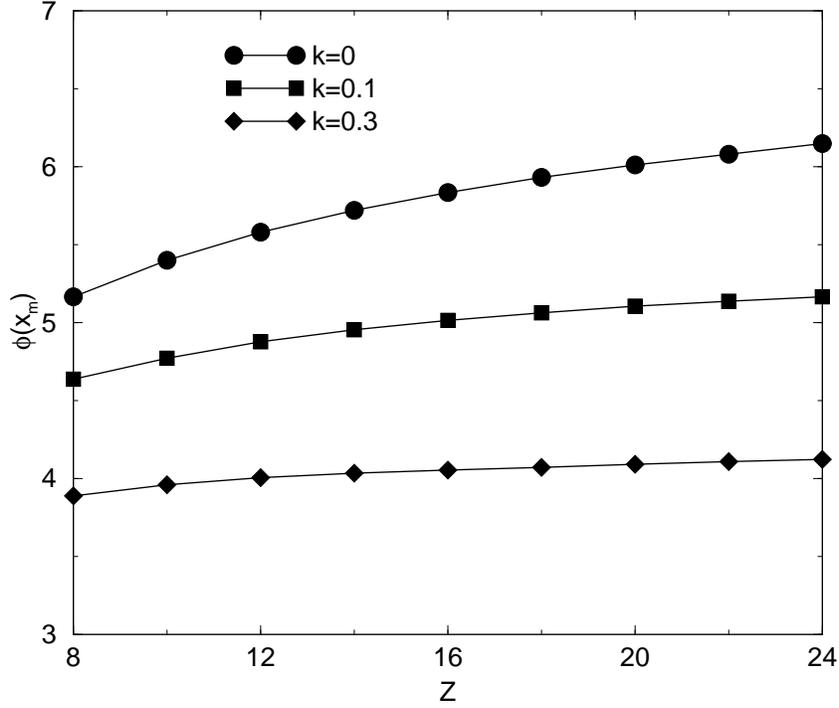}
\caption{The barrier for polyions with $\xi=1$ and 
$n^*$ condensed trivalent counterions, such that  $\phi_{n^*}(0)=0$,  
as a function of $Z$. 
}
\label{Dphiz}
\end{figure}
\end{center}

We next explore the dependence of the  
barrier height $\phi(x_m)$, 
where $x_m$ is the position of the maximum of 
the potential of mean force 
see Fig. \ref{phix},  on the parameters of the model.
In Fig. \ref{Dphixi410},  $\phi(x_m)$ 
is depicted as a function of the Manning
parameter $\xi$ for polyion of size $Z=10$ with $n=4$ 
associated counterions.
We see that the barrier height
diminishes with the increase of $\xi$ 
and the amount of salt inside the suspension. 
To explore the dependence of the barrier height
on the size of the polyion $Z$, in Fig. \ref{Dphiz}
we plot $\phi(x_m)$ as a function of $Z$ for complexes composed
of a polyion and 
$n^*$ condensed trivalent counterions, such that  $\phi_{n^*}(0)=0$.
While in the absence of salt the barrier height shows a significant
dependence on the polyion size, at finite salt
concentration this dependence weakens and $\phi(x_m)$ seems to saturate 
when the polyion size is significantly larger than the Debye length. For
large $Z$ and small concentration of electrolyte, however, 
the kinetic barrier  
can be many $k_B T$, providing a significant limitation
to overcharging~\cite{LeAr03a}. 

\begin{center}
\begin{figure}
\includegraphics[scale=0.7]{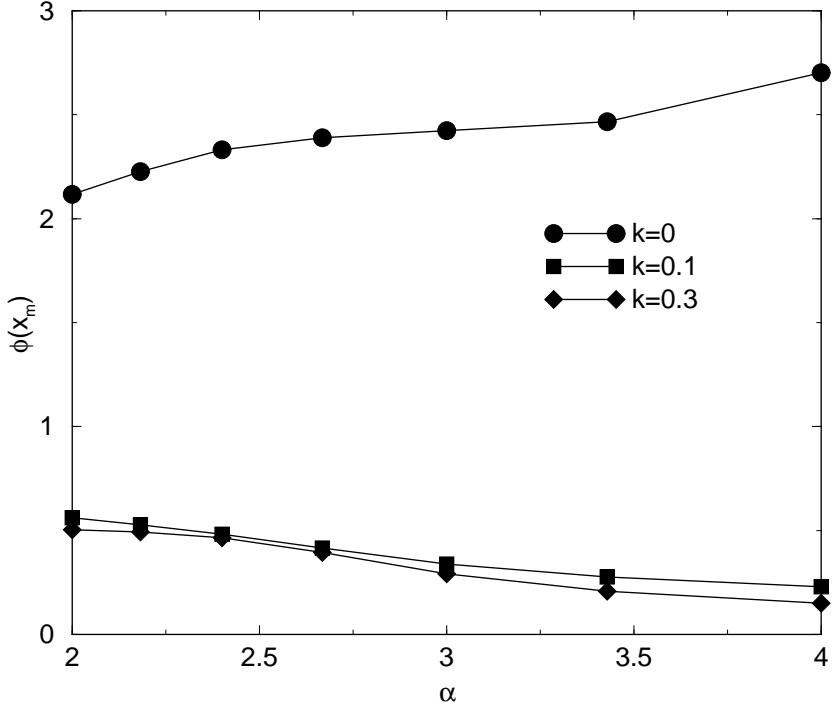}
\caption{The barrier for complexes
composed of $Z=20$ and $n$ condensed counterions,
such that $n\alpha=24$, as a function of the
counterion valence $\alpha$. }
\label{Dphialpha}
\end{figure}
\end{center}

\begin{center}
\begin{figure}
\includegraphics[scale=0.7]{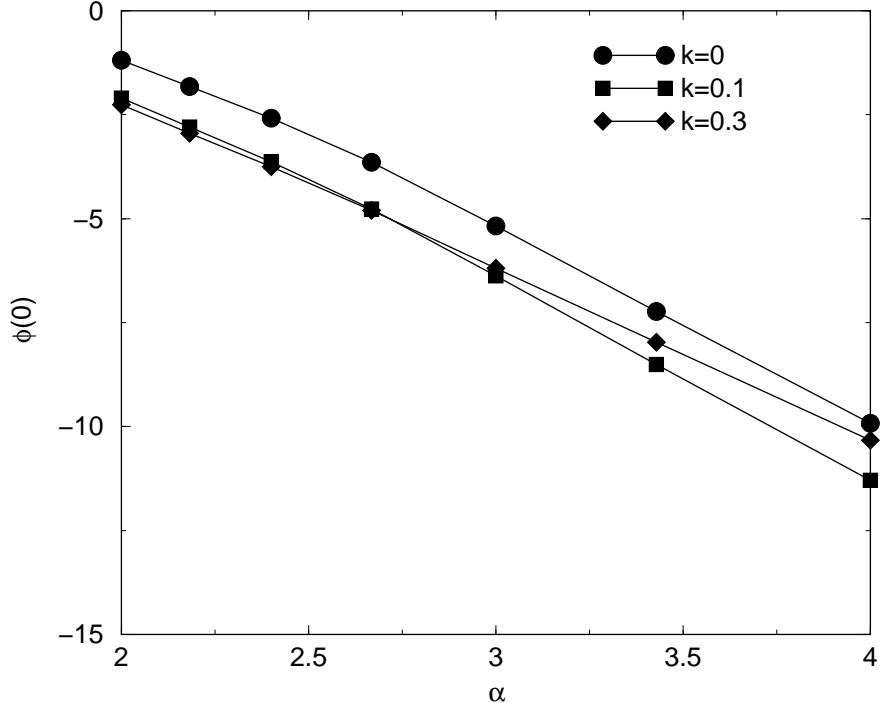}
\caption{$\phi(0)$ for complexes
composed of $Z=20$ and $n$ condensed counterions,
such that $n\alpha=24$, as a function of the
counterion valence $\alpha$. }
\label{dphialpha}
\end{figure}
\end{center}

Charge reversal is a consequence of strong positional correlations
between the counterions.  These correlations
are induced by the electrostatic repulsion between the particles. Thus,
we expect that both the barrier height and the relative depth
of the absolute minimum will be strongly dependent on the counterion 
valence.  In Figs. \ref{Dphialpha} and 
\ref{dphialpha}  we show the dependence
of the barrier height and the depth of the potential well 
on the valence of the counterions.
Although all the overcharged complexes 
depicted in Figs. \ref{Dphialpha} and 
\ref{dphialpha} have the same net
charge $4q$, the depth of the potential well
and the height of the kinetic barrier depend on $\alpha$. 
As expected, larger counterion charge
leads to stronger positional correlations and 
favors the counterion condensation and the charge reversal ($\phi(0)$
becomes more negative with increasing $\alpha$). The barrier height,
however, once again shows a nontrivial dependence on the 
salt concentration.  For small amounts of salt and large $Z$,
increased counterion valence leads to larger kinetic barriers.

\section{Conclusions}

We have studied the potential of mean force between
a polyion and an $\alpha$-valent counterion
inside a polyelectrolyte solution 
containing multivalent counterions and monovalent
salt.  The model is sufficiently simple that the partition function
can be calculated exactly.  It is found that for an overcharged polyion  
the potential of mean force can have two minima,  one located at $x=0$
and another $x=\infty$.  Which one of the minima is the global
one depends on the charge density of the polyion and the
amount of salt inside the suspension.  When the global minimum is
at $x=0$, a counterion from the bulk
finds it energetically favorable to approach
the polyion surface. To reach $x=0$, however, the counterion must
overcome a free energy barrier.  For small
salt concentrations, this barrier can be sufficiently large
to provide a kinetic limitation to the extent of charge reversal.
Furthermore, even if the counterion reaches $x=0$, whether
or not it will become  condensed will depend on the
depth of the potential well.  Counterion condensation
will occur only if  $\phi(x_m)-\phi(0) \gg 1/\xi$.  Otherwise,
the thermal fluctuations will lead to a fast escape of the
counterion from the $x=0$ minimum. 

For suspensions containing rodlike polyelectrolytes and the
multivalent counterions micro phase separation  is observed
under certain conditions~\cite{TaJa96,Br01}.  
The polyions aggregate
forming bundles
with a well defined crossectional area.  
It has been argued that bundle formation is an activated
process and the size of the bundles is kinetically 
controlled~\cite{Sh99,HaLi99,StLeAr02}.
It should then be quite interesting to explore the dependence
of the barrier height on the concentration of monovalent
electrolyte using a theory similar to the one presented above.

To conclude, the extent of the charge reversal is  
strongly dependent on the
amount of monovalent salt present in the suspension.  
Small concentrations of salt will enhance the overcharging 
while an excessive amount of salt will hinder the charge reversal.
Furthermore, even if the minimum of the free energy corresponds
to an overcharged state, we find that depending on the polyion
charge density and the amount of salt in the suspension,
there can be significant kinetic limitations to the overcharging.

This work was supported in part by the Brazilian agencies
CNPq and FAPERJ. 
JFS acknowledges funding by project Pronex-CNPq-FAPERJ/171.168-2003.


\begin{thebibliography}{32}
\expandafter\ifx\csname natexlab\endcsname\relax\def\natexlab#1{#1}\fi
\expandafter\ifx\csname bibnamefont\endcsname\relax
  \def\bibnamefont#1{#1}\fi
\expandafter\ifx\csname bibfnamefont\endcsname\relax
  \def\bibfnamefont#1{#1}\fi
\expandafter\ifx\csname citenamefont\endcsname\relax
  \def\citenamefont#1{#1}\fi
\expandafter\ifx\csname url\endcsname\relax
  \def\url#1{\texttt{#1}}\fi
\expandafter\ifx\csname urlprefix\endcsname\relax\def\urlprefix{URL }\fi
\providecommand{\bibinfo}[2]{#2}
\providecommand{\eprint}[2][]{\url{#2}}

\bibitem[{\citenamefont{Widom}(1965)}]{Wi65}
\bibinfo{author}{\bibfnamefont{B.}~\bibnamefont{Widom}}, \bibinfo{journal}{J.
  Chem. Phys.} \textbf{\bibinfo{volume}{43}}, \bibinfo{pages}{3898}
  (\bibinfo{year}{1965}).

\bibitem[{\citenamefont{Widom}(1984)}]{Wi84}
\bibinfo{author}{\bibfnamefont{B.}~\bibnamefont{Widom}}, \bibinfo{journal}{J.
  Chem. Phys.} \textbf{\bibinfo{volume}{81}}, \bibinfo{pages}{1030}
  (\bibinfo{year}{1984}).

\bibitem[{\citenamefont{Kolomeisky and Widom}(1999)}]{Wi99}
\bibinfo{author}{\bibfnamefont{A.~B.} \bibnamefont{Kolomeisky}}
  \bibnamefont{and} \bibinfo{author}{\bibfnamefont{B.}~\bibnamefont{Widom}},
  \bibinfo{journal}{Faraday Discussions} \textbf{\bibinfo{volume}{112}},
  \bibinfo{pages}{81} (\bibinfo{year}{1999}).

\bibitem[{\citenamefont{Levin}(2002)}]{Le02}
\bibinfo{author}{\bibfnamefont{Y.}~\bibnamefont{Levin}}, \bibinfo{journal}{Rep.
  Prog. Phys.} \textbf{\bibinfo{volume}{65}}, \bibinfo{pages}{1577}
  (\bibinfo{year}{2002}).

\bibitem[{\citenamefont{Luijten et~al.}(2002)\citenamefont{Luijten, Fisher, and
  Panagiotopoulos}}]{LuFiPa01}
\bibinfo{author}{\bibfnamefont{E.}~\bibnamefont{Luijten}},
  \bibinfo{author}{\bibfnamefont{M.~E.} \bibnamefont{Fisher}},
  \bibnamefont{and} \bibinfo{author}{\bibfnamefont{A.~Z.}
  \bibnamefont{Panagiotopoulos}}, \bibinfo{journal}{Phys. Rev. Lett.}
  \textbf{\bibinfo{volume}{88}}, \bibinfo{pages}{185701}
  (\bibinfo{year}{2002}).

\bibitem[{\citenamefont{van Roij and Hansen}(1997)}]{RoHa97}
\bibinfo{author}{\bibfnamefont{R.}~\bibnamefont{van Roij}} \bibnamefont{and}
  \bibinfo{author}{\bibfnamefont{J.~P.} \bibnamefont{Hansen}},
  \bibinfo{journal}{Phys. Rev. Lett.} \textbf{\bibinfo{volume}{79}},
  \bibinfo{pages}{3082} (\bibinfo{year}{1997}).

\bibitem[{\citenamefont{Levin et~al.}(1998)\citenamefont{Levin, Barbosa, and
  Tamashiro}}]{LeBaTa98}
\bibinfo{author}{\bibfnamefont{Y.}~\bibnamefont{Levin}},
  \bibinfo{author}{\bibfnamefont{M.~C.} \bibnamefont{Barbosa}},
  \bibnamefont{and} \bibinfo{author}{\bibfnamefont{M.~N.}
  \bibnamefont{Tamashiro}}, \bibinfo{journal}{Europhys. Lett.}
  \textbf{\bibinfo{volume}{41}}, \bibinfo{pages}{123} (\bibinfo{year}{1998}).

\bibitem[{\citenamefont{Diehl et~al.}(2001)\citenamefont{Diehl, Barbosa, and
  Levin}}]{DiBaLe01}
\bibinfo{author}{\bibfnamefont{A.}~\bibnamefont{Diehl}},
  \bibinfo{author}{\bibfnamefont{M.~C.} \bibnamefont{Barbosa}},
  \bibnamefont{and} \bibinfo{author}{\bibfnamefont{Y.}~\bibnamefont{Levin}},
  \bibinfo{journal}{Europhys. Lett.} \textbf{\bibinfo{volume}{53}},
  \bibinfo{pages}{86} (\bibinfo{year}{2001}).

\bibitem[{\citenamefont{Deserno and Gr\"unberg}(2002)}]{DeGr02}
\bibinfo{author}{\bibfnamefont{M.}~\bibnamefont{Deserno}} \bibnamefont{and}
  \bibinfo{author}{\bibfnamefont{H.~H.} \bibnamefont{Gr\"unberg}},
  \bibinfo{journal}{Phys. Rev. E} \textbf{\bibinfo{volume}{66}},
  \bibinfo{pages}{011401} (\bibinfo{year}{2002}).

\bibitem[{\citenamefont{Tamashiro and Schiessel}(2003)}]{TaSc03}
\bibinfo{author}{\bibfnamefont{M.~N.} \bibnamefont{Tamashiro}}
  \bibnamefont{and}
  \bibinfo{author}{\bibfnamefont{H.}~\bibnamefont{Schiessel}},
  \bibinfo{journal}{J. Chem. Phys.} \textbf{\bibinfo{volume}{119}},
  \bibinfo{pages}{1855} (\bibinfo{year}{2003}).

\bibitem[{\citenamefont{Trizac and Levin}(2004)}]{TrLe04}
\bibinfo{author}{\bibfnamefont{E.}~\bibnamefont{Trizac}} \bibnamefont{and}
  \bibinfo{author}{\bibfnamefont{Y.}~\bibnamefont{Levin}},
  \bibinfo{journal}{Phys. Rev. E} \textbf{\bibinfo{volume}{69}},
  \bibinfo{pages}{031403} (\bibinfo{year}{2004}).

\bibitem[{\citenamefont{Rouzina and Bloomfield}(1996)}]{RoBl96}
\bibinfo{author}{\bibfnamefont{I.}~\bibnamefont{Rouzina}} \bibnamefont{and}
  \bibinfo{author}{\bibfnamefont{V.}~\bibnamefont{Bloomfield}},
  \bibinfo{journal}{J. Chem. Phys.} \textbf{\bibinfo{volume}{100}},
  \bibinfo{pages}{9977} (\bibinfo{year}{1996}).

\bibitem[{\citenamefont{Gr\o{}nbech-Jensen
  et~al.}(1997)\citenamefont{Gr\o{}nbech-Jensen, Mashl, Bruinsma, and
  Gelbart}}]{GrMaBr97}
\bibinfo{author}{\bibfnamefont{N.}~\bibnamefont{Gr\o{}nbech-Jensen}},
  \bibinfo{author}{\bibfnamefont{R.~J.} \bibnamefont{Mashl}},
  \bibinfo{author}{\bibfnamefont{R.~F.} \bibnamefont{Bruinsma}},
  \bibnamefont{and} \bibinfo{author}{\bibfnamefont{W.~M.}
  \bibnamefont{Gelbart}}, \bibinfo{journal}{Phys. Rev. Lett.}
  \textbf{\bibinfo{volume}{78}}, \bibinfo{pages}{2477} (\bibinfo{year}{1997}).

\bibitem[{\citenamefont{Ha and Liu}(1997)}]{HaLi97}
\bibinfo{author}{\bibfnamefont{B.-Y.} \bibnamefont{Ha}} \bibnamefont{and}
  \bibinfo{author}{\bibfnamefont{A.~J.} \bibnamefont{Liu}},
  \bibinfo{journal}{Phys. Rev. Lett.} \textbf{\bibinfo{volume}{79}},
  \bibinfo{pages}{1289} (\bibinfo{year}{1997}).

\bibitem[{\citenamefont{Arenzon et~al.}(1999)\citenamefont{Arenzon, Stilck, and
  Levin}}]{ArStLe99}
\bibinfo{author}{\bibfnamefont{J.~J.} \bibnamefont{Arenzon}},
  \bibinfo{author}{\bibfnamefont{J.~F.} \bibnamefont{Stilck}},
  \bibnamefont{and} \bibinfo{author}{\bibfnamefont{Y.}~\bibnamefont{Levin}},
  \bibinfo{journal}{Eur. Phys. J. B} \textbf{\bibinfo{volume}{12}},
  \bibinfo{pages}{79} (\bibinfo{year}{1999}).

\bibitem[{\citenamefont{Hansen and L\"owen}(2000)}]{HaLo00}
\bibinfo{author}{\bibfnamefont{J.~P.} \bibnamefont{Hansen}} \bibnamefont{and}
  \bibinfo{author}{\bibfnamefont{H.}~\bibnamefont{L\"owen}},
  \bibinfo{journal}{Annual Reviews of Phys. Chem,}
  \textbf{\bibinfo{volume}{51}}, \bibinfo{pages}{209} (\bibinfo{year}{2000}).

\bibitem[{\citenamefont{Gelbart et~al.}(2000)\citenamefont{Gelbart, Bruinsma,
  Pincus, and Parsegian}}]{GeBrPi00}
\bibinfo{author}{\bibfnamefont{W.~M.} \bibnamefont{Gelbart}},
  \bibinfo{author}{\bibfnamefont{R.~F.} \bibnamefont{Bruinsma}},
  \bibinfo{author}{\bibfnamefont{P.~A.} \bibnamefont{Pincus}},
  \bibnamefont{and} \bibinfo{author}{\bibfnamefont{V.~A.}
  \bibnamefont{Parsegian}}, \bibinfo{journal}{Physics Today}
  \textbf{\bibinfo{volume}{53}}, \bibinfo{pages}{38} (\bibinfo{year}{2000}).

\bibitem[{\citenamefont{Solis and de~la Cruz}(2001)}]{SoCr01}
\bibinfo{author}{\bibfnamefont{F.~J.} \bibnamefont{Solis}} \bibnamefont{and}
  \bibinfo{author}{\bibfnamefont{M.~O.} \bibnamefont{de~la Cruz}},
  \bibinfo{journal}{Physics Today} \textbf{\bibinfo{volume}{54}},
  \bibinfo{pages}{71} (\bibinfo{year}{2001}).

\bibitem[{\citenamefont{Angelini et~al.}(2003)\citenamefont{Angelini, Liang,
  Wrigglers, and Wong}}]{AnLiWr03}
\bibinfo{author}{\bibfnamefont{T.~E.} \bibnamefont{Angelini}},
  \bibinfo{author}{\bibfnamefont{H.}~\bibnamefont{Liang}},
  \bibinfo{author}{\bibfnamefont{W.}~\bibnamefont{Wrigglers}},
  \bibnamefont{and} \bibinfo{author}{\bibfnamefont{G.~C.~L.}
  \bibnamefont{Wong}}, \bibinfo{journal}{PNAS} \textbf{\bibinfo{volume}{100}},
  \bibinfo{pages}{8634} (\bibinfo{year}{2003}).

\bibitem[{\citenamefont{Lozada-Cassou et~al.}(1982)\citenamefont{Lozada-Cassou,
  Saavedra-Barrera, and Henderson}}]{LoSaHe82}
\bibinfo{author}{\bibfnamefont{M.}~\bibnamefont{Lozada-Cassou}},
  \bibinfo{author}{\bibfnamefont{R.}~\bibnamefont{Saavedra-Barrera}},
  \bibnamefont{and}
  \bibinfo{author}{\bibfnamefont{D.}~\bibnamefont{Henderson}},
  \bibinfo{journal}{J. Chem. Phys.} \textbf{\bibinfo{volume}{77}},
  \bibinfo{pages}{5150} (\bibinfo{year}{1982}).

\bibitem[{\citenamefont{Messina et~al.}(2002)\citenamefont{Messina, Tovar,
  Lozada-Cassou, and Holm}}]{MeToLo01}
\bibinfo{author}{\bibfnamefont{R.}~\bibnamefont{Messina}},
  \bibinfo{author}{\bibfnamefont{E.~G.} \bibnamefont{Tovar}},
  \bibinfo{author}{\bibfnamefont{M.}~\bibnamefont{Lozada-Cassou}},
  \bibnamefont{and} \bibinfo{author}{\bibfnamefont{C.}~\bibnamefont{Holm}},
  \bibinfo{journal}{Europhys. Lett.} \textbf{\bibinfo{volume}{60}},
  \bibinfo{pages}{383} (\bibinfo{year}{2002}).

\bibitem[{\citenamefont{Grosberg et~al.}(2002)\citenamefont{Grosberg, Nguyen,
  and Shklovskii}}]{GrNgSh02}
\bibinfo{author}{\bibfnamefont{A.~Y.} \bibnamefont{Grosberg}},
  \bibinfo{author}{\bibfnamefont{T.~T.} \bibnamefont{Nguyen}},
  \bibnamefont{and} \bibinfo{author}{\bibfnamefont{B.~I.}
  \bibnamefont{Shklovskii}}, \bibinfo{journal}{Rev. Mod. Phys.}
  \textbf{\bibinfo{volume}{74}}, \bibinfo{pages}{329} (\bibinfo{year}{2002}).

\bibitem[{\citenamefont{Martin-Molina et~al.}(2003)\citenamefont{Martin-Molina,
  Quesada-Perez, Galisteo-Gonzalez, and Hidalgo-Alvarez}}]{MaQuGa03}
\bibinfo{author}{\bibfnamefont{A.}~\bibnamefont{Martin-Molina}},
  \bibinfo{author}{\bibfnamefont{M.}~\bibnamefont{Quesada-Perez}},
  \bibinfo{author}{\bibfnamefont{F.}~\bibnamefont{Galisteo-Gonzalez}},
  \bibnamefont{and}
  \bibinfo{author}{\bibfnamefont{R.}~\bibnamefont{Hidalgo-Alvarez}},
  \bibinfo{journal}{J. Phys.: Condensed Mat.} \textbf{\bibinfo{volume}{15}},
  \bibinfo{pages}{S3475} (\bibinfo{year}{2003}).

\bibitem[{\citenamefont{Debye and H\"uckel}(1923)}]{DeHu23}
\bibinfo{author}{\bibfnamefont{P.~W.} \bibnamefont{Debye}} \bibnamefont{and}
  \bibinfo{author}{\bibfnamefont{E.}~\bibnamefont{H\"uckel}},
  \bibinfo{journal}{Phys. Z.} \textbf{\bibinfo{volume}{24}},
  \bibinfo{pages}{185} (\bibinfo{year}{1923}).

\bibitem[{\citenamefont{Manning}(1969)}]{Ma69}
\bibinfo{author}{\bibfnamefont{G.~S.} \bibnamefont{Manning}},
  \bibinfo{journal}{J. Chem. Phys.} \textbf{\bibinfo{volume}{51}},
  \bibinfo{pages}{924} (\bibinfo{year}{1969}).

\bibitem[{\citenamefont{Manning}(1978)}]{Ma78}
\bibinfo{author}{\bibfnamefont{G.~S.} \bibnamefont{Manning}},
  \bibinfo{journal}{Q. Rev. Biophys. II} \textbf{\bibinfo{volume}{2}},
  \bibinfo{pages}{179} (\bibinfo{year}{1978}).

\bibitem[{\citenamefont{Levin and Arenzon}(2003)}]{LeAr03a}
\bibinfo{author}{\bibfnamefont{Y.}~\bibnamefont{Levin}} \bibnamefont{and}
  \bibinfo{author}{\bibfnamefont{J.~J.} \bibnamefont{Arenzon}},
  \bibinfo{journal}{J. Phys. A: Math. Gen} \textbf{\bibinfo{volume}{36}},
  \bibinfo{pages}{5857} (\bibinfo{year}{2003}).

\bibitem[{\citenamefont{Tang and Janmey}(1996)}]{TaJa96}
\bibinfo{author}{\bibfnamefont{J.~X.} \bibnamefont{Tang}} \bibnamefont{and}
  \bibinfo{author}{\bibfnamefont{P.~A.} \bibnamefont{Janmey}},
  \bibinfo{journal}{J. Biol. Chem.} \textbf{\bibinfo{volume}{271}},
  \bibinfo{pages}{8556} (\bibinfo{year}{1996}).

\bibitem[{\citenamefont{Bruinsma}(2001)}]{Br01}
\bibinfo{author}{\bibfnamefont{R.}~\bibnamefont{Bruinsma}},
  \bibinfo{journal}{Phys. Rev. E} \textbf{\bibinfo{volume}{63}},
  \bibinfo{pages}{061705} (\bibinfo{year}{2001}).

\bibitem[{\citenamefont{Shklovskii}(1999)}]{Sh99}
\bibinfo{author}{\bibfnamefont{B.~I.} \bibnamefont{Shklovskii}},
  \bibinfo{journal}{Phys.Rev.Lett.} \textbf{\bibinfo{volume}{82}},
  \bibinfo{pages}{3628} (\bibinfo{year}{1999}).

\bibitem[{\citenamefont{Ha and Liu}(1999)}]{HaLi99}
\bibinfo{author}{\bibfnamefont{B.-Y.} \bibnamefont{Ha}} \bibnamefont{and}
  \bibinfo{author}{\bibfnamefont{A.~J.} \bibnamefont{Liu}},
  \bibinfo{journal}{Europhys. Lett.} \textbf{\bibinfo{volume}{46}},
  \bibinfo{pages}{624} (\bibinfo{year}{1999}).

\bibitem[{\citenamefont{Stilck et~al.}(2002)\citenamefont{Stilck, Levin, and
  Arenzon}}]{StLeAr02}
\bibinfo{author}{\bibfnamefont{J.~F.} \bibnamefont{Stilck}},
  \bibinfo{author}{\bibfnamefont{Y.}~\bibnamefont{Levin}}, \bibnamefont{and}
  \bibinfo{author}{\bibfnamefont{J.~J.} \bibnamefont{Arenzon}},
  \bibinfo{journal}{J. Stat. Phys.} \textbf{\bibinfo{volume}{106}},
  \bibinfo{pages}{287} (\bibinfo{year}{2002}).

\end{thebibliography}

\end{document}